# Optical Evidence for Mixed Phase Behavior in Manganites

P. Gao[1], T. A. Tyson[1], Z. Liu[2], M. A. DeLeon[1] and C. Dubourdieu[3]

[1]*Department of Physics, New Jersey Institute of Technology, Newark, NJ 07102*

[2]*Geophysical Laboratory, Carnegie Institution of Washington, Washington, D.C., 20015*

[3]*Laboratoire des Matériaux et du Génie Physique, CNRS, Grenoble INP, 3 parvis Louis Néel, BP 257, 38016 Grenoble, France*

## Abstract

Synchrotron infrared measurements were conducted over the range 100 to 8000 cm$^{-1}$ on a self-doped La$_x$MnO$_{3-\delta}$ ($x\sim0.8$) film. From these measurements we determined the conductivity, the effective number of free carriers, and the specific phonon modes as a function of frequency. While the metal-insulator transition temperature ($T_{MI}$) and the magnetic ordering temperature ($T_C$) approximately coincide, the free carrier density onset occurs at a significantly lower temperature (~45 K below). This suggests that local distortions exist below $T_{MI}$ and $T_C$ which trap the $e_g$ conduction electrons. These regions with local distortions constitute an insulating phase which persists for temperatures significantly below $T_{MI}$ and $T_C$. The initial large drop in resistivity is due to the enhanced magnetic ordering while further drops correspond to reductions in the insulating phase which increase the number of free carriers.

PACS Numbers: **78.20.-e, 72.80.-r, 75.47.Lx, 64.75.St**

$R_{1-x}A_xMnO_3$ (R: trivalent rear-earth ions, A: divalent alkaline-earth ions) has been widely studied due to the large magnetoresistance observed [1 2 3 4 5 6]. Divalent cation doping induces a change from $Mn^{3+}$ to $Mn^{4+}$. The induced holes in the $e_g$ level create a mixed valence system. These materials have also attracted much theoretical and fundamental physics interest since they exhibit intimate coupling of spin, lattice, orbital, and charge degrees of freedom. This coupling results in a ground state energy landscape with multiple minima corresponding to different charge, spin and structural configurations. Small external perturbations (such as temperature, pressure, substrate strain, magnetic fields, and electric fields) can shift the system from one state to another. A mixed valence on Mn sites can also be induced in the La-deficient $La_{1-x}MnO_{3-\delta}$ (x>0, δ>0) manganite. Both ferromagnetic order and metallic conductivity can be obtained [7 8 9] and the transition temperatures can be adjusted by both the oxygen content and the La deficiency [10 11].

One of the open questions about these colossal magnetoresistive oxides concerns the transition of the system from the high temperature paramagnetic insulating phase to the low temperature conductive ferromagnetic phase [12 13]. It is thought that just above the metal to insulator transition, regions of metallic phase begin to grow within the insulating host and then dominate at low temperature based on structural and optical mode measurements [14 15]. The nature of this mixed phase behavior is still under discussion. While a qualitative model of the structural changes exists, an understanding of the parallel changes in spin and free carrier numbers has not been achieved. Optical experiments are a good way to study the phonon modes and electron-phonon coupling in oxides [16]. In addition, information on the free carrier concentration can be derived.

In this work, we report on a synchrotron infrared spectroscopic study on $La_{0.8}MnO_{3-\delta}$ films deposited on a $LaAlO_3$ substrate. Reflection measurements were carried out over a broad temperature range including the metal-insulator transition region and fits were performed to extract the optical conductivity and free electron carrier numbers at each temperature.

Self-doped $La_xMnO_{3-\delta}$ (x~0.8) films were epitaxially grown on (001) $LaAlO_3$ (LAO) substrates by liquid injection metal organic chemical vapor deposition. The detailed procedure is described elsewhere [10]. Here we report the results for a ~120 nm thick film which were also found in a ~410 nm film. X-ray diffraction measurement shows that such thick films have low strain [17]. The *in situ* post-deposition annealing leads indeed to the strain relaxation. The magnetization was measured between 5 and 400 K range under a 0.2 T magnetic field. The film resistance was measured by a 4-point probe setup in the temperature range of 10 to 320 K. As seen in Fig. 1 the metal-insulator phase transition temperature (282 K) and the Curie temperature (275.2 K) are quite close and a high magnetization saturation of 3.56 $\mu_B$/Mn is achieved. The resistivity at low temperature (10 K) has a value of $4.6 \times 10^{-4}$ Ωcm, which compares to (5 K) $2 \times 10^{-4}$ Ωcm for $La_{1-x}Ca_xMnO_3$ (x=0.33) films on a LAO substrate [18].

Synchrotron reflectivity spectra were measured at U2A beamline at the National Synchrotron Light Source, Brookhaven National Laboratory. This beamline has a Bruker IFS 66v/S vacuum spectrometer equipped with a Bruker IRscope-II microscope, a MCT detector, and a KBr beamsplitter for mid-IR; a custom made infrared microscope with long working distant (40 mm) reflecting objective, a 3.5-micron mylar beamsplitter and a Si bolometer detector for far-IR. The infrared frequency range covers 100-8000 cm$^{-1}$ and a spectral resolution of 4 cm$^{-1}$ is applied to all spectra. A ~0.5 $\mu m$ thick gold layer was deposited on the top of the La$_{0.8}$MnO$_{3-\delta}$ film as a reference mirror for the reflection measurements. The sample was mounted on the cold finger of a continuous flow cryostat and the measurement temperatures were 304 (Start at 304 K), 282, 275, 265, 255, 245, 225, 200, 150, 125, 100, 80, 50, 20, and 10 K.

The reflectivity for the film and the bare substrate are given in Figs. 2(a) and (b), respectively. Note the systematic enhancement of the reflectivity with decreasing temperature. The LAO substrate is an insulator and the small spectra variations at different temperatures can be seen in Fig. 2(b). In Fig. 2(a), the blue lines, which are spectra in the vicinity of T$_{MI}$ and T$_C$ (304 K, 282 K, and 275 K) show LAO-like resonances in the frequency range 100 to 550 cm$^{-1}$ and small differences are found in the region above 550 cm$^{-1}$. The reflectivity peak near ~180 cm$^{-1}$ is close to 1 because of the interference.

The reflectivity spectra on an expanded scale (100 to 550 cm$^{-1}$) are plotted in Fig. 2(c). As the temperature decreases (between 275 K and 200 K), the film resistance drops and the film becomes more reflective. Free carriers screen the substrate, and as a result, the broad peaks at ~200 cm$^{-1}$ and ~450 cm$^{-1}$ become narrow and the reflectivity drops. This phenomenon could be the direct evidence of incomplete conversion to the metallic phase. Between 150 K and 10 K, the reflectivity increases gradually in the whole frequency range except around the peaks at ~200 cm$^{-1}$ and ~450 cm$^{-1}$. A saturation level is approached below 200 K. More information can be obtained by examining the details of the spectra. In the reflectivity spectra, phonon modes can be seen as resonances at low frequency.

When the penetration depth [19] is of the order of the film thickness, the IR beam can pass through the 120 nm FM metallic film and reach the substrate particularly at low frequency. Modeling a one-side polished LAO substrate (0.5 mm thick) as a half-infinite plate, the total reflectivity depends on the film thickness and the dielectric functions [20]. Using the Drude-Lorentz model [19]:

$$\varepsilon(\omega) = \varepsilon_\infty - \frac{\omega_p^2}{\omega(\omega + i\Gamma)} + \sum_j \frac{\omega_{p_j}^2}{(\omega_{o_j}^2 - \omega^2) - i\omega\Gamma_j}$$ for both the LAO substrate [20] and the La$_{0.8}$MnO$_{3-\delta}$ film,

we fit the reflectivity spectra at all temperatures and a fitting spectrum at 10 K are shown in Fig. 3(a). For the 10 K fit, 13 components (8 for film and 5 for substrate) were used. Each component had unique $\omega_{oj}$, $\Gamma_j$ and $\omega_{pj}$ parameters. From the fit, the resonance frequencies, amplitudes and decay lifetimes can be extracted. The optical conductivity can be obtained from the dielectric function: $\sigma(\omega) = -\frac{i\omega\varepsilon(\omega)}{4\pi}$.

Temperature dependent optical conductivity spectra as a function of frequency are plotted in Fig. 3 for the film (Fig. 3(b)) and the substrate (Fig. 3(c)). The weight of the Drude components of the film increases as the temperature decreases and the peak above 1000 cm$^{-1}$ which has been assigned to a small polaron [21 22] moves toward the low frequency side with decreasing temperature. In the vicinity of T$_{MI}$ and T$_C$, the optical conductivity spectra are similar to each other and the polaron peak shifts are small. The peak maximum shifts from ~5200 cm$^{-1}$ at 304 K to ~4000 cm$^{-1}$ at 265 K. The polaron shifts to the low energy side quickly and the spectral weight increases with reduced temperature. Below 200 K, the reflectivity spectra and the optical spectra vary slowly with the temperature. The Drude free carriers contribution dominates the conductivity spectra. Fig. 3(c) shows LAO substrate optical spectra at two different temperatures. The positions do not change significantly with temperature (< 4 cm$^{-1}$) while the amplitudes vary.

Photoionization of a polaron system has been shown to exhibit a distinct frequency dependence, which depends on the degree of localization of the final electron state [22]. For excitations involving transitions between highly localized states (adjacent Mn sites), the absorption profile is quite symmetric. This contrasts with excitations to free-electron like states which produce an absorption curve with a sharp low-energy rise and a long high-energy tail. The transitions between localized states are associated with so-called small polarons. As can be seen in Fig. 4, the polaron resonance is highly symmetric.

In addition, we can calculate the resistivity from the plasma frequency $\omega_p$ in the Drude model. The zero frequency conductivity, plasma frequency and resistivity are related by: $\sigma(0) = \frac{ne^2\tau}{m}$, $\omega_p^2 = \frac{4\pi ne^2}{m}$, $\rho = \frac{1}{4\pi\omega_p^2\tau}$. The DC resistivity and calculated resistivity (from the zero frequency conductivity) are shown together in Fig. 5(a). The plot shows a good agreement between the two independent data sets, indicating that the chosen model is appropriate. They have a consistent trend and show a good agreement for temperatures where the sample is more metallic.

Fig. 5(b) and 5(c) display the vibration modes near ~580 cm$^{-1}$ (in-plane oxygen) and near ~630 cm$^{-1}$ [23] as a function of temperature. We can combine the information obtained from the temperature dependence of resistivity and the vibrational modes. When the temperature is lower than $T_C$, the Jahn-Teller distortion decreases and the charge carriers are delocalized (as seen in the decrease in the resistivity curve). The distortion persists until the temperature is below ~234 K ($T_{onset}$ defined as the inflection point in Fig. 5(d)). At lower temperatures, as the distortions are reduced, the oxygen vibrational modes should harden in the higher symmetry configuration. It can be seen in the temperature variation of the vibrational frequencies. These stretching mode shifts are quite close with the observations by Kim [24], Boris [25] and Hartinger [26].

The effective carrier number density $N_{eff}(T, \omega_c)$ is proportional to the integrated optical conductivity spectral weight.

$$N_{eff}(T,\omega_c) = \frac{2mV}{\pi e^2} \int_0^{\omega_c} \text{Re}\,\sigma(\omega')d\omega'$$

We use a cut-off frequency ($\omega_c$) of 5000 cm$^{-1}$ to calculate the number density and $N_{eff}(T)$ is plotted in Fig. 5(d). By examining the cut-off frequency dependence of the shape of $N_{eff}$ [19 27] vs. temperature, we found that the onset profile is independent of $\omega_c$ for cut-off frequencies ranging from 4800 to 32,000 cm$^{-1}$.

It can be seen that the onset of saturation in carrier density (~234 K) is significantly below $T_{MI}$ and $T_C$ (275 K).

The results suggest a more complex mechanism for the transition to the low temperature phase in manganites. It is consistent with the existence of regions with significant local distortions below $T_{MI}$ and $T_C$ which trap the $e_g$ conduction electrons. These regions with local distortions constitute an insulating phase which persists for temperatures significantly below $T_{MI}$ and $T_C$. Low spin scattering will lead to the observed initial large resistivity drop [28] as a result of magnetic ordering of the $t_{2g}$ spins enabling Mn–Mn site hopping [29 30] when reducing temperature below $T_C$. Further reductions in resistivity are then due to reductions in the volume of the minority insulating phase which then increases the number of free carriers. The origin of the flattening of the reflectivity spectra at low temperature is due to the increased number of free carriers, which limits the penetration depth of the light into the sample.

In conclusion, we have explored the temperature dependent infrared reflectivity spectra of $La_xMnO_{3-\delta}$ 120 nm /LAO ($x$~0.8) system over the range 100 to 8000 cm$^{-1}$. A polaron resonance in the mid-infrared region is observed. The conductivity was found to be systematically enhanced at lower temperatures. While the metal-insulator transition temperature ($T_{MI}$) and the magnetic ordering temperature ($T_C$) approximately coincide, the free carrier density onset occurs at a significantly lower temperature (~45 K below). These results are consistent with the existence of insulating regions at temperatures significantly below the metal insulator and magnetic ordering temperatures.

This research was funded by NSF DMR-0512196, NSF INT-0233316, and CNRS/NSF project No. 14550. U2A beamline is supported by COMPRES, the Consortium for Materials Properties Research in Earth Sciences under NSF Cooperative Agreement EAR01-35554, U.S. Department of Energy (DOE-BES and NNSA/CDAC). Use of the National Synchrotron Light Source, Brookhaven National Laboratory, was supported by the U.S. Department of Energy, Office of Science, Office of Basic Energy Sciences, under Contract No. DE-AC02-98CH10886.

# Figure Captions

**Fig. 1.** Resistivity and magnetization of $La_{0.8}MnO_{3-\delta}$/LAO film.

**Fig. 2.** (a) $La_{0.8}MnO_{3-\delta}$/LAO reflectivity spectra at different temperatures: 10, 20, 50, 80, 100, 125, 150 (purple), 200 (green), 225, 245, 255, 265, 275, 282, 304 K on a logarithmic energy scale. The vertical scale corresponds to the 304 K spectrum, and the others are shifted up by 0.1 relative to the previous temperature. (b) LAO substrate reflectivity spectra on a logarithmic energy scale. (c) Expansion of curves in (a) over the range 100 to 550 $cm^{-1}$ on a linear energy scale.

**Fig. 3.** (a)An example fit of $La_{0.8}MnO_{3-\delta}$/LAO reflectivity spectra at 10 K. (b) Real part of the optical conductivity spectra of $La_{0.8}MnO_{3-\delta}$ at different temperatures (purple and green lines corresponding to 150 K and 200 K) (b) Real part of the optical conductivity spectra of LAO substrate.

**Fig. 4.** Real part of the optical conductivity of $La_{0.8}MnO_{3-\delta}$ at different temperatures on a linear energy scale.

**Fig. 5.** (a) Magnetization, calculated and measured DC resistivity. (b) In-plane oxygen vibrational mode and (c) higher frequency mode. (d) Effective free carrier number.

**Fig. 1. (Gao *et al.*)**

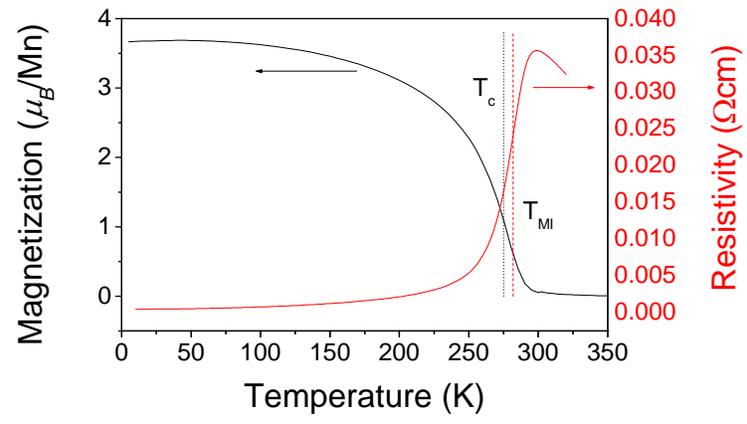

**Fig.2. (Gao *et al*.)**

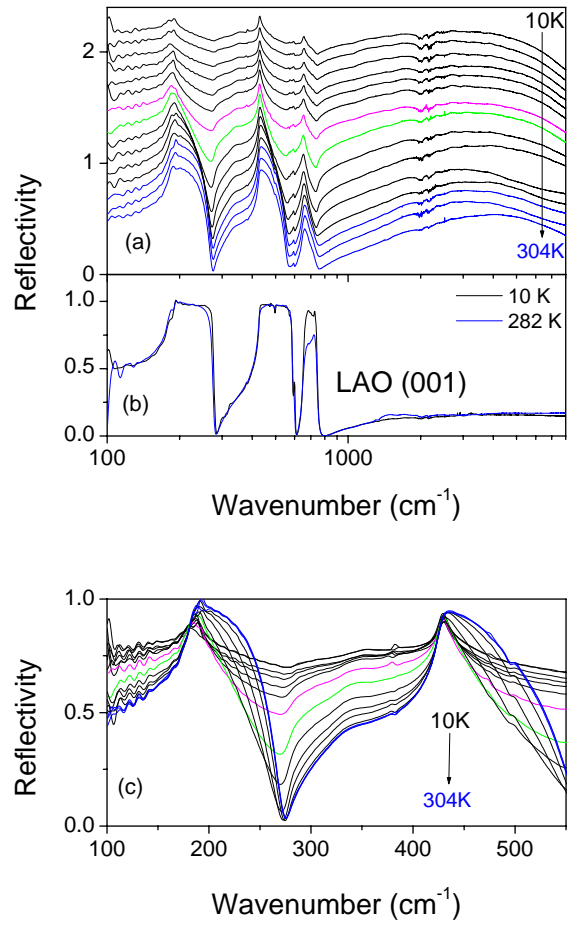

**Fig. 3. (Gao *et al.*)**

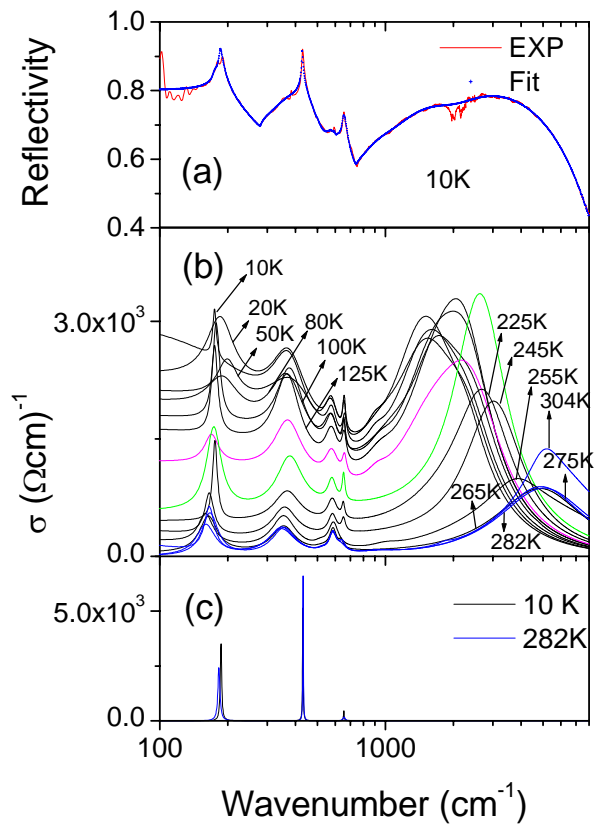

**Fig. 4. (Gao *et al.*)**

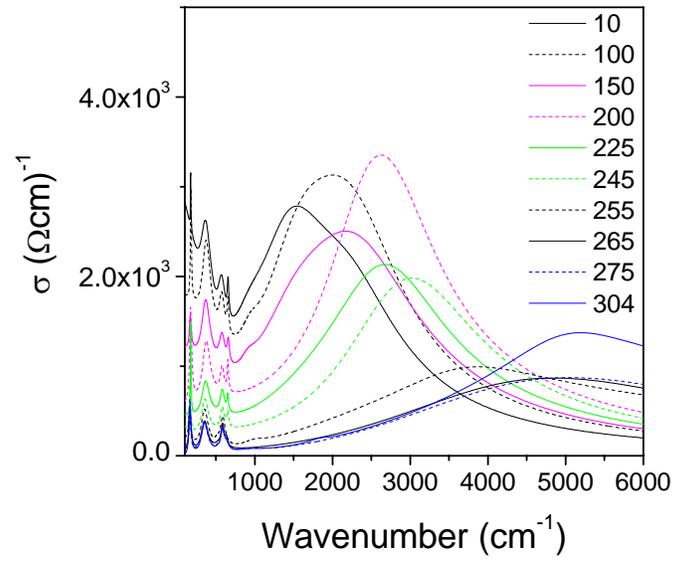

**Fig. 5. (Gao *et al.*)**

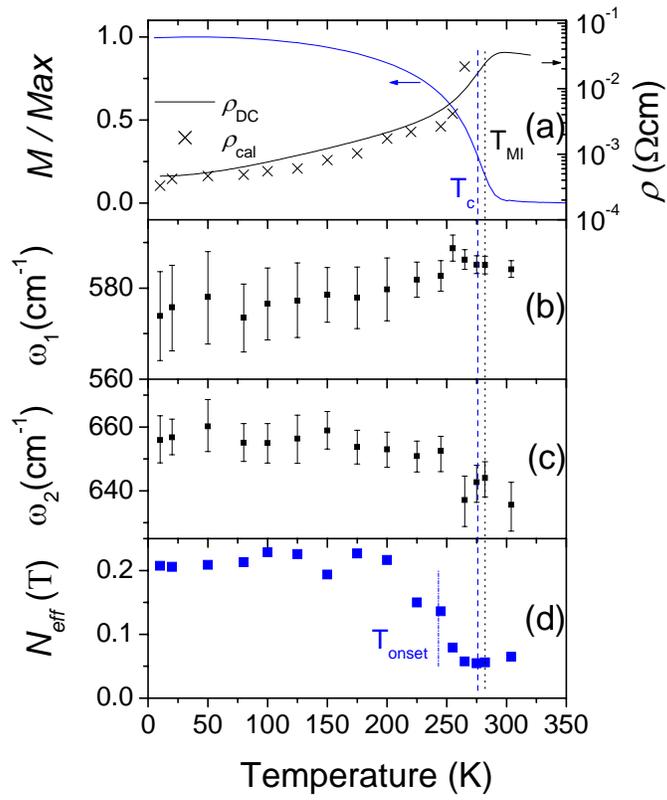

[1] S. Jin, T. H. Tiefel, M. McCormack, R. A. Fastnacht, R. Ramesh, and L. H. Chen, Science 264, 413 (1994).
[2] C. N. R. Rao, A. K. Cheetham, and R. Mahesh, Chem. Mater. 8 (10), 2421 (1996).
[3] M. B. Salamon and M. Jaime, Rev. Mod. Phys. 73 (3), 583 (2001).
[4] J. M. D. Coey, M. Viret, and S. Von Molnar, Adv. Phys. 48 (2), 167 (1999).
[5] Y. Tokura (Ed.), Colossal Magnetoresistive Oxides, Gordon and Breach Science, London (1999).
[6] W. Prellier, Ph Lecoeur, and B. Mercey, J. Phys.: Condens. Matter 13 (48), 915 (2001).
[7] J. Topfer and J. B. Goodenough, Chem. Mater. 9 (6), 1467 (1997).
[8] A. Maignan, C. Michel, M. Hervieu, and B. Raveau, Solid State Commun. 101 (4), 277 (1997).
[9] A. A. Bosak, O. Yu Gorbenko, A. R. Kaul, I. E. Graboy, C. Dubourdieu, J. P. Senateur, and H. W. Zandbergen, J. Magn. Magn. Mater. 211 (1-3), 61 (2000).
[10] A. Bosak, C. Dubourdieu, M. Audier, J. P. Sénateur, and J. Pierre, Appl. Phys. A, 79 (8), 1979 (2004).
[11] Z. Chen, T. A. Tyson, K. H. Ahn, Z. Zhong, and J. Hu, arXiv: 0708. 1963.
[12] A. Moreo, S. Yunoki, and E. Dagotto, Science 283 (5410), 2034 (1999).
[13] K. H. Ahn, T. Lookman, and A. R. Bishop, Nature 428 (6981), 401 (2004).
[14] S. J. L. Billinge, T. Proffen, V. Petkov, L. J. Sarrao, and S. Kycia, Phys. Rev. B 62 (2), 1203 (2000).
[15] H. L. Liu, S. Yoon, S. L. Cooper, S. W. Cheong, P. D. Han, and D. A. Payne, Phys. Rev. B 58 (16), 10115 (1998).
[16] F. Gervais, Mater. Sci. Eng. R: Reports, 39 (2-3), 29 (2002).
[17] Q. Qian, T. A. Tyson, C. Dubourdieu , A. Bossak, J. P. Senateur, M. Deleon, J. Bai, G. Bonfait, and J. Maria, J. Appl. Phys. 92 (8), 4518 (2002).
[18] G. Jeffrey Snyder, R. Hiskes, S. DiCarolis, M. R. Beasley, and T. H. Geballe, Phys. Rev. B 53 (21), 14434 (1996).
[19] F. P. Mena, A. B. Kuzmenko, A. Hadipour, J. L. M. van Mechelen, D. van der Marel, and N. A. Babushkina, Phys. Rev. B 72 (13), 134422 (2005)
[20] Z. M. Zhang, B. I. Choi, M. I. Flik, and A. C. Anderson, J. Opt. Soc. Am. B 11 (11), 2252 (1994).
[21] Ch. Hartinger, F. Mayr, A. Loidl, and T. Kopp, Phys. Rev. B 73 (2), 024408 (2006).
[22] D. Emin, Phys. Rev. B 48 (18), 13691 (1993).
[23] I. Fedorov, J. Lorenzana, P. Dore, G. De Marzi, P. Maselli, P. Calvani, S. W. Cheong, S. Koval, and R. Migoni, Phys. Rev. B 60 (17), 11875 (1999).
[24] K. H. Kim, J. Y. Gu, H. S. Choi, G. W. Park, and T. W. Noh, Phys. Rev. Lett. 77 (9), 1877 (1996).
[25] A. V. Boris, N. N. Kovaleva, A. V. Bazhenov, P. J. M. van Bentum, Th. Rasing, S. W. Cheong, A. V. Samoilov, and N. C. Yeh, Phys. Rev. B 59 (2), 697 (1999).
[26] Ch. Hartinger, F. Mayr, A. Loidl, and T. Kopp, Phys. Rev. B 70 (13), 134415 (2004).
[27] M. Quijada, J. Černe, J. R. Simpson, H. D. Drew, K. H. Ahn, A. J. Millis, R. Shreekala, R. Ramesh, M. Rajeswari, and T. Venkatesan, Phys. Rev. B 58 (24), 16093 (1998).
[28] Y. Lyanda-Geller, S. H. Chun, M. B. Salamon, P. M. Goldbart, P. D. Han, Y. Tomioka, A. Asamitsu, and Y. Tokura, Phys. Rev. B 63 (18), 184426 (2001).
[29] Y. Tomioka, H. Kuwahara, A. Asamitsu, T. Kimura, R. Kumai, and Y. Tokura, Physica B: Condens. Matter 246-247, 135 (1998).
[30] Y. Tomioka, A. Asamitsu, H. Kuwahara, Y. Moritomo, and Y. Tokura, Phys. Rev. B 53 (4), 1689 (1996).